**Evolution in complex objects**

Mourad Oussalah

Nantes University

*2, rue de la Houssinière - BP 92208*
*44322 Nantes Cedex 03 France*
*e-mail : Mourad.Oussalah@univ-nantes.fr*

**Abstract**

This paper describes work carried out on a model for the evolution of graph classes in complex objects. By defining evolution rules and propagation strategies on graph classes, we aim to define a user-definable means to manage data evolution model which tackles the complex nature of the classes managed, using the concepts defined in object systems.
So, depending on their needs and on those of the targeted application, designers can choose the evolution mechanism they consider to suit them best. They can either create new evolutions or reuse predefined ones to respond to a given need.

## 1. Introduction

Today's industrial application of components, services and architectures technology has to take the data evolution problem into consideration. Indeed, the data structures dealt with by the engineering applications like CAD applications, telecommunication networks, mechanics, architecture, are more and more complex. We are interested in supporting evolution in a liberal rather then a conservative fashion; rather than the system offering a list of possible and fixed evolutions to the designer, we think that the designer should be able to specify adapted rules to his needs and rely on the system for assistance and verification. Our approach differs from the one usually chosen by evolution models [1, 2, 3] which provide fixed evolution management. It consists in giving the possibility to the designer to dynamically define an evolution which takes into account of the different semantics of the connected classes in a given application.

In this work we are essentially interested by the evolution of graph classes. We consider a graph class as a semantic graph composed of nodes classes and semantic relations classes(inheritance relation, part-whole relation, association relation,…) which can themselves describe a graph class and so on. The evolution consists either in permitting the **changes** [7] of the structure of the graph class or in permitting its **versioning** [6]. In the first case, the evolution concerns the changes occurring on the graph/node/relation classes and their potential mutual impact. The changes on a graph/node/relation class concern the class itself (adding and deletion of the class, modification of the name of the class), its definition (adding or deletion of an attribute, modification of the name of an attribute, adding or deletion of a method, modification of the name of a method).
In the second case, the evolution concerns the versioning of a graph/node/relation class via creation, deletion and propagation operations of version of theses classes.

In this paper, we propose a model for evolution including both changes of the structure and versioning of graph classes [8, 9], as well as a dynamic management of their modification. To each kind of class (graph, node, relation) is associated a propagation strategy able to define explicitly its evolution police. The aim is to have a tool making it possible to select an element of the graph class to be modified, to carry out the modification and to propagate modification through the graph by creating new classes versions if necessary. The tool will also give the possibility to modify propagation strategies and their corresponding evolution rules and to reuse them.

## 2. Motivating factor and Objectives

The main motivating factor is to maintain in a uniform way the consistency of the evolution of a graph class by permitting the changes of its structure and its versioning and by respecting its semantic. This consistency is achieved via a perturbation model : starting from a graph class which is initially consistent, an element of this class evolves (node, semantic relation, attribute,…) and the task of the system is to find back a new consistent graph class.
Many applications require the use of graph classes (inheritance graph, composition graph,…) and their evolution [2]. So, we have defined objectives to be reached for our graph class evolution model :

- An abstraction level of the evolution must be provided; this abstraction level will allow evolutions to be reusable and generic.
- Evolution must be managed outside the classes concerned by the evolutions; indeed, merging the evolution behaviour and the methods which describe the behaviour specific to each class runs counter to the behaviour abstraction.
- The evolution model must be open to the addition of new external methods of evolution .
- The evolution model must be able to take into account the semantics of various types of relation of a graph class and not impose fixed evolution police.
- The evolution model must be able to take advantage of the features of the object-oriented paradigm such as abstraction, polymorphism, encapsulation, etc. More precisely, the principle of reusability must be widely exploited. To begin with, an evolution can concern several distinct sets of classes. Moreover, a new evolution can be defined by combining evolutions which have already been defined (via the inheritance or composition relations).
- The use of evolution must be flexible and easy to manage.

Having considered these objectives, we have chosen to represent an evolution of a graph class in two distinct components :

Evolution = propagation strategy component + evolution rule component and to keep the relationship (graph Class and its evolution police) in a component called evolution manager. This last aspect means that the principle of encapsulation is not violated, and evolutions concerning a given class are easily attained from the class asked to evolve. In what follows, we will present the different aspects of the proposed evolution model.

## 2. Basic concepts

In order to better understand the concepts introduced in this paper, it is necessary to define the basic concepts we rely on :

First of all, the concept of *graph class*, which is the support of our modelling, is a semantic graph composed of nodes classes and semantic relations classes like inheritance, composition or association relations. These semantic relations specify precisely the quality of existing interactions between nodes or graphs. In many models, the composition relation, for example, conveys strong semantics [1, 6]: a composition relation can, for instance, be exclusive or shared, dependent or independent, predominant or not. Moreover, relations can have a direction. A relation can be an afferent one or efferent to a node. In our model each kind of class (graph or node or relation) is reified and then owns its structure and its behaviour and in this case its evolution.

In order to express this evolution, the designer is able to attach evolution capabilities directly to his applications classes concerned by the evolution; of course he can also , by default, keep the evolution police provided by the system. Indeed, in our model, the evolution of a class is based on two components : propagation strategies and evolution rules. A Propagation strategy groups together the set of evolution rules which define the operations of creation, destruction, modification, derivation, versioning applicable to a given class (graph, node or relation). A propagation strategy, if it exists, is therefore associated with each class graph or node or relation, it can be reused or redefined in the corresponding sub-classes hierarchies. An evolution rules defines declaratively the actions that must be triggered on the classes concerned by the evolution. The evolution rules are defined as active rules and are reified so they can be hierarchical [5] : they are based on the formalism of ECA rules (Event/ Condition/ Action) and are hierarchical via the inheritance relation.
For example, the version creation or the version destruction rules of a node via the Action part of its evolution rule will trigger the evolution rules of the corresponding afferent and efferent relations associated with the processed node. For the relations, these rules can be propagated in four directions and according to two modes. The propagation *direction* of a relation evolution rule can be FORWARD, BACKWARD, BIDIRECTIONNEL or NONE. FORWARD, for example, means that the propagation takes place from the source of the relation to its destination. The propagation *mode* can be RESTRICTED or EXTENDED. If it is RESTRICTED, the operation propagates from the extremity on which it is triggered to the relation class. If it is EXTENDED, the operation propagates from the extremity on which it is triggered to both the relation class and the other extremity of the relation.

The use of propagation strategies containing evolution rules allows a great flexibility because rules can be defined and carried out according to the contexts and needs of an information system. Otherwise, in order to simplify the use of evolution rules classes and propagation strategies classes, we defined predefined ones adapted to the most current used in object oriented databases systems [8].

## 3. Evolution model architecture

The architecture of our graph class evolution model is based on the following key concepts (cf. figure 1) : modelled graph class, evolution manager, propagation strategies and evolution rules :

By modelled *graph class* we mean a *graph class composed of nodes class and semantic relations class*. All these classes have all in common to own a structure, a behaviour and an evolution police(changes and versioning). So, *Meta-Class* and the *graph/node/relation* (in figure 1) are defined at meta level. These meta-classes can also be specialized. Then, an evolution concerning a class is defined at meta level and is applied at class level. It can be defined at class level and then is applied at instance level. In this paper, we will detail only evolution described at meta level and applied at class level.
The applications classes which interest our study are defined at the class level and their evolution are ruled by the meta level.

The proposed evolution model is based on three fundamental notions :
The first is the concept of *evolution manager* which is responsible of the interception of an evolution message (create-class, destroy-class, create-version class,…) received by a given class, then it accesses the propagation strategy which manages the evolution of a class and triggers the evolution rules corresponding to the event, if they exist. The triggering of rules results in the execution of their action part.

The evolution manager groups together the information necessary for the management and propagation of all the evolutions which concern a given class. It allows a class to have direct access to the evolutions which concern it. The evolution manager is considered as an abstract meta-class.

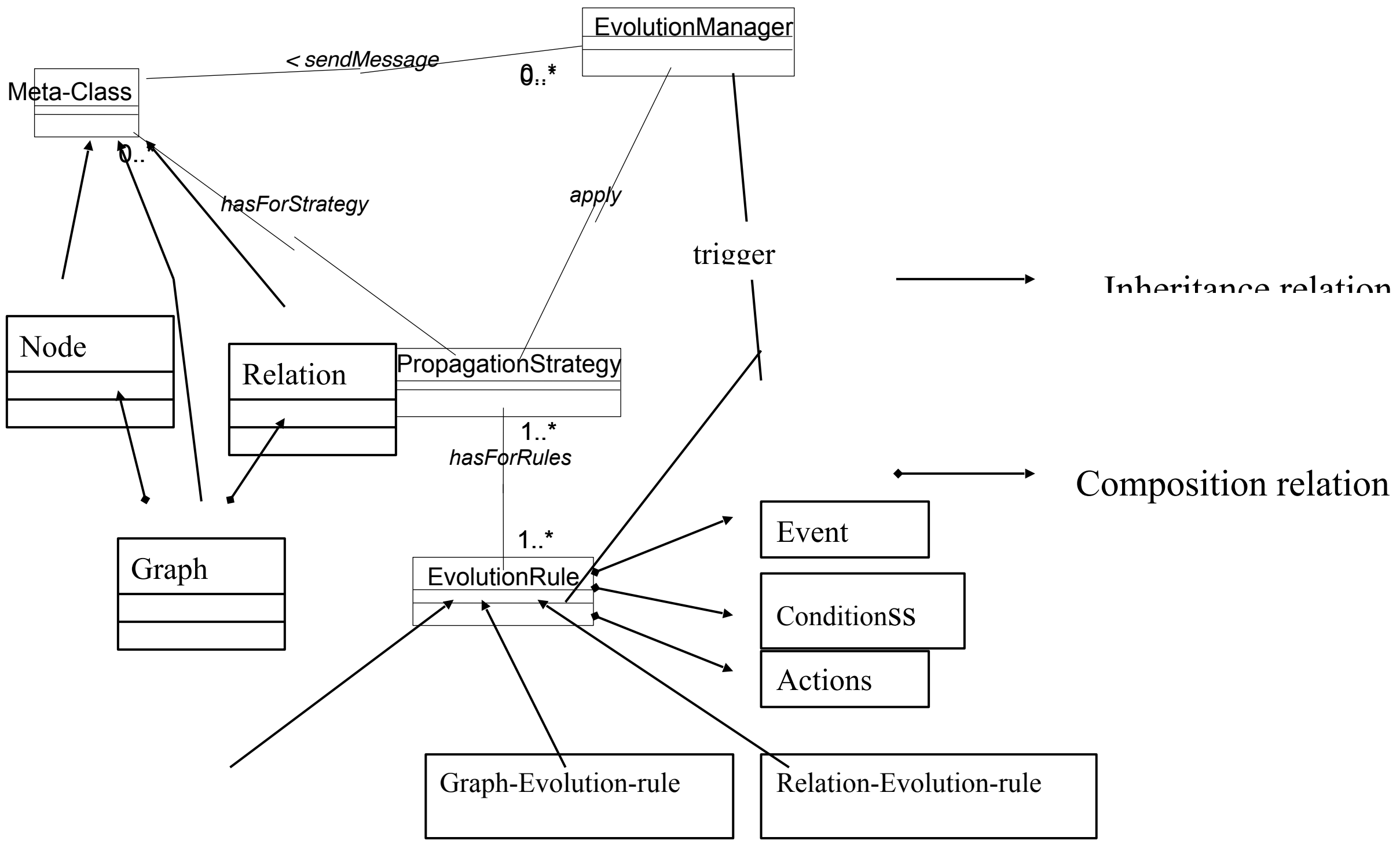

Figure 1: Evolution model architecture at the meta-level

The second corresponds to the concept *propagation strategy* which is composed of a group of evolution rules associated with the graph/node/relation meta-classes. The third concept corresponds to the concept *Evolution rule*. An *evolution rule* is composed of an event, conditions and actions. When an event occurs, if the condition holds, then the corresponding action is executed. This action can concern both the evolution rule on the class itself and its eventual impact on the others classes. We consider three kind of evolution rules in a graph class : There are specific evolution rules applying for the graphs, nodes and relations evolution. Moreover, when it concerns a semantic relation, its properties must be taken into account. Indeed, a semantic relation is directed, has a propagation direction and a mode (restrictive or extend). In the case of the restrictive mode, the evolution rule acts only on the considered class, whereas in the extend mode it also acts on the classes attached to the other extremity. These properties are significant because they define a propagation behaviour of a relation.

The meta-classes *Propagation Strategy* and *Evolution rule* are generic and can be instantiated. The corresponding instances which are themselves classes can be specialized and reused but are considered as abstract classes.

## 4. Evolution operating mechanism

The operating mechanism describes the execution process of the evolution model. It is composed of three steps.

**Step 1 : Interception of the event to execute**

An evolution can come from two different ways :

- after a user request.
  Indeed, the user selects both the class of type (graph, node, relation) concerned by the evolution and the rule to apply to it (deletion, modification, versioning…). The evolution manager intercepts the message representing the user choice.

- after the execution of an evolution rule action part.
  Indeed, the execution of an action part of an evolution rule can involve the call of an other event, and so on until the propagation is over. So, the evolution manager is responsible of the interception of any new event.

**Step 2 : Research of propagation strategies and execution of evolution rules**

The evolution manager having received a request of an evolution of a class, then looks for the corresponding propagation strategy (if it exists) and then applies this strategy to the class and triggers the corresponding evolution rules.

These rules are identified by the event type to execute (for example for a node evolution the corresponding event is : delete-node, create-node-version, delete-attribute-node,…) and are applied after the condition are checked. Actions of these rules can be a program code or eventually a list of events to be executed on other classes.

**Step 3 : Propagation**

The triggering of evolution rules results in the execution of their action part. This execution can raise new events that will be executed in the same way , and recursively propagate other evolution rules.

In order to avoid cycles in the activation of rules, the evolution manager stores the names of classes that have been treated during a given propagation process. This prevents messages concerning the same class from being taken into account more than once.

## 5. Example

The example of the figure 3 allows to illustrate the evolution of the graph class GR0 (figure 3)

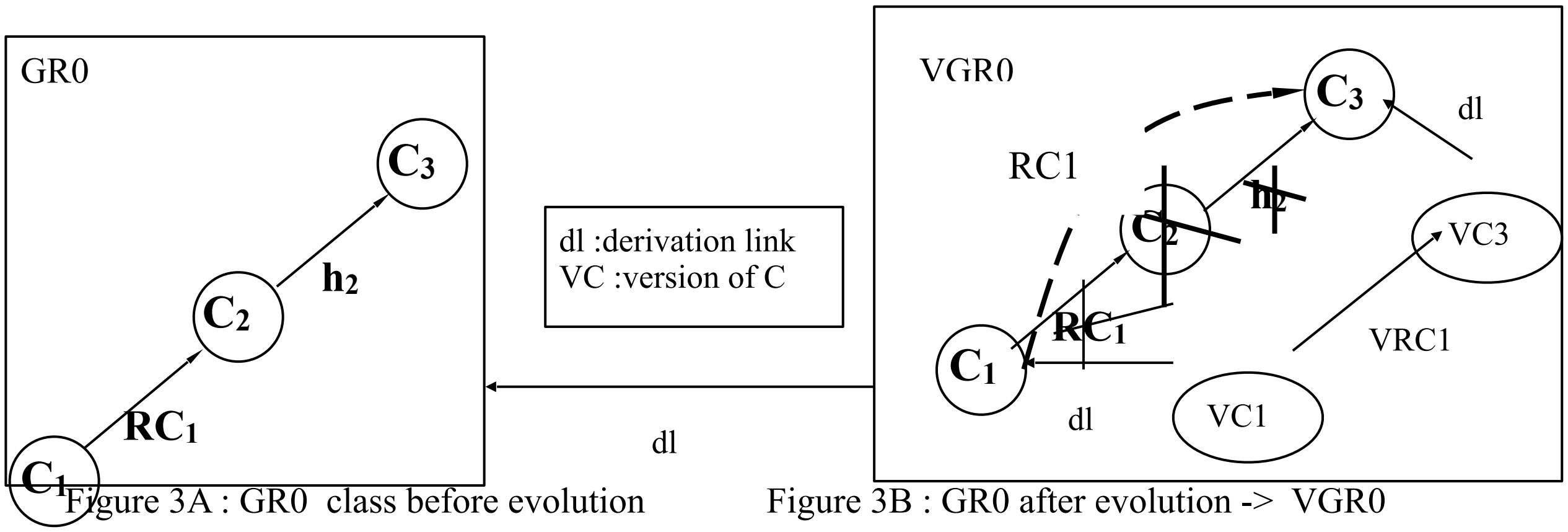

Figure 3A : GR0 class before evolution    Figure 3B : GR0 after evolution -> VGR0

Scenario : the user selects the C2 class and decides first to delete it (figure 3A) and then to create a version of the C1 class. The results of this evolution (figure 3B) depend on the different evolutions described below by the designer.

The different classes acting in this evolution are :

| $C_1$ : Node | $C_2$ : Node | $C_3$ : Node |
|---|---|---|
| ***afferente relations :*** - | ***afferente relations:*** $RC_1$ | ***afferente relations:*** $h_2$ |
| ***efferente relations :*** $RC_1$ | ***efferente relations:*** $h_2$ | ***efferente relations:*** - |
| ***attributes :*** … | ***attributes :*** … | ***attributes :*** … |
| ***methods :*** … | ***methods :*** … | ***methods :*** … |

| $RC_1$ : Relation | | $h_2$ : Relation | | GR0 : Graph |
|---|---|---|---|---|
| ***nature :*** | composition | ***nature :*** | inheritance | ***Nodes : C1, C2, C3*** |
| ***source :*** | $C_1$ | ***source :*** | $C_2$ | ***Relations : RC1, h2*** |
| ***destination :*** | $C_2$ | ***destination :*** | $C_3$ | |
| ***exclusive :*** | true | ***exclusive :*** | true | |
| ***dependent :*** | false | ***dependent :*** | false | |
| ***predominent :*** | false | ***predominent :*** | false | |
| ***card :*** | 1 | ***card :*** | 1 | |
| ***reverse-card :*** | 1 | ***reverse-card:*** | 1 | |

| PropagationStrategy | S1: Graph | S2: Node | S3 : Relation |
|---|---|---|---|
| ***isTheDefaultStrategyForClass :*** | ***GR0*** | ***C1,C2,C3*** | ***RC1,h2*** |
| ***hasAsCreationRules :*** | ***R9*** | ***R7*** | ***R1,R8*** |
| ***hasAsDestructionRules :*** | | ***R2*** | ***R4*** |
| ***HasAsmodifucationRules :*** | ***R3,R5*** | ***R6*** | |

| $R_1$ : RelationEvolutionRule | $R_2$ : NodeEvolutionRule | $R_3$ : GraphEvolutionRule |
|---|---|---|
| **Event** : addrelation(R,N1,N2,G)<br><br>**Condition** :<br>Belong(N1,G)<br>Belong(N2,G)<br><br>**Actions** :<br>Instantiaterelation(R,N1,N2,G) | ***event :*** deleteNode(N)<br><br>***condition :*** not shared(N)<br>G <- Graph(N)<br><br>***actions :*** modifygraph(G,N,())<br>executedeletenode(N) | ***event :*** modifygraph(G,N,())<br><br>***condition :*** belong(N,G)<br>R1<- N.afferent<br>R2<- N.efferent<br><br>***actions:***<br>deleterelation(R1)<br>modifnode-(R1.source,efferent,R1)<br>deleterelation(R2)<br>modifnode-(R2.destination,afferent,R2)<br>G.relations <-G.relations – {R1,R2}<br>addRelation(R1.name,R1.source,R2.destination, G)<br><br>G.Node<-G.Node – N<br><br>G.relations<- G.relations + R1 |

| $R_4$ : RelationEvolutionRule | R5: GraphEvolutionRule | R6: NodeEvolutionRule |
|---|---|---|
| ***direction :*** forward<br><br>***mode :*** extended<br><br>***event :*** deleterelation(R)<br><br>***Condition:*** G<- Graph(R)<br><br>***actions :***<br>modifygraph(G,R,())<br>executedeleterelation(R) | ***Event :*** modifygraph(G,R,())<br><br>***Condition :*** belong(R,G)<br>N1<-R.source<br>N2<-R.Destination<br><br>***Actions :***<br>Modifnode-(N1,efferent,R)<br><br>Modifnode-(N2,afferent,R) | ***Event :*** modifnode-(N,type,R)<br><br>***Condition :***<br>Belong(R, N.afferent) or<br>Belong(R, N.efferent)<br><br>***Actions :***<br>Case type of<br>Afferent : N.afferent<-N.afferent - R<br>Efferent : N.efferent<-N.efferent - R |

| R7 : NodeEvolutionRule | R8 : RelationEvolutionRule | R9 : GraphEvolutionRule |
|---|---|---|
| **Event :** create-version-node(N) | **Direction :** forward<br>Mode : extended | **Event** : create-version-graph(G,N) |
| **Conditions :** versionable (N) | **Event :**<br>create-version-relation(r,N,N1)<br>**conditions :**<br>V(N) exists | **Conditions :**<br>Belong(N,G)<br>let r(N,N1) and r.RelationOperationRule.mode = extended |
| **Actions :**<br>V(N)<- execute-create-version(N)<br>G <- graph(N)<br>Create-version-graph(G,N) | **Actions :**<br>V(r) <- derive(r)<br>V(N1) <- create-version-node(N1)<br>V(r).source <- V(N)<br>V(r).destination <- V(N1) | **Actions :**<br>Create-version-relation(r,N,N1)<br>V(G)<-execute-create-version(G) |

**Steps of operations after the deletion of the node $C_2$ and the creation of the version of C1 node**

The deletion of the $C_2$ class consists not only in deleting it, but also in propagating (following the propagation strategies) the deletion to the other classes which depend on it, like the composition relation $RC_1$ and the inheritance relation $h_2$. The propagation of this modification is managed by the $S_2$ propagation strategy and more precisely by its destruction R2 rule. Indeed, the evolution manager applies the strategy $S_2$ which consists in bringing back its operation rule $R_2$ dealing with the deletion of a node and then triggers it. The description of the rule $R_2$ consists ,before deleting the node C2, to verify the conditions of this deletion (the afferent and efferent relations of the node C2 must be exclusive), and then in executing the actions "modifygraph(G,N,())" and "executedeleteNode(N)". So, the evolution manager intercepts the next event consisting in : "modifygraph(G,N,())". This event is send to the graph entity GR0 to which we have associated the strategy S1 which owns two modification rules R3 and R5. In this case, the rule R3 is selected by the evolution manager. The other operations follow these steps :

- Strategy S1 , rule R3 on graph GR0 //modification of the graph GR0 after the deletion of the node C2
  - Strategy S3, rule R4 on relation RC1 //deletion of the relation RC1
    - Strategy S1, rule R5 on graph GR0// modification of the graph GR0
      - Strategy S2, rule R6 on node C1//modification of the node C1
      - Strategy S2, rule R6 on node C2//modification of the node C2
  - Strategy S2, rule R6 on node C1// modification of the node C1, if-needed
  - Strategy S3, rule R4 on relation h2//deletion of the relation h2
    - Strategy S1, rule R5 on graph GR0//modification of the graph GR0
    - Strategy S2, rule R6 on node C2//modification of the node C2, if-needed
    - Strategy S2, rule R6 on node C3//modification of the node C3
  - Strategy S2, rule R6 on node C3// modification of the node C3, if-needed
  - Strategy S3, rule R1 on relation RC1//creation of the relation RC1

Concerning the creation of the version of the C1 node, the following rules are triggered :

- Strategy S2, rule R7 on node C1// creation of the version VC1
  - Strategy S1, rule R9 on graph GR0//update graph GR0
    - Strategy S3, rule R8 on relation RC1//creation of the version of RC1 relation :VRC1
      //creation of the version of the C3 node : VC3
      //connecting VC1 and VC3 via VRC1
  - Strategy S1, rule R9 on graph GR0// creation of the version of GR0 graph : VGR0

By default, the new creating classes (VRC1,VC1,VC3 and VGR0) are associated to predefined strategies and rules of classes types which they depend. However, the designer is free to redefine or specialize them for a targeted application.

## Conclusion

The proposed evolution model respects most of the objectives we determined before the design process. In addition to the mechanisms which are inherent in the representation of evolutions (propagation strategies and evolution rules) by objects of first class, the specialization of evolutions and application graph classes may be dealt with independently. In this way, a graph class is simpler and easier to read, and evolutions are more easily expressed by the designer. The principal originality of our model lies in the fact that different semantics of graph class evolution can be taken into account.
Moreover, it differs from the existing models in two points :

- it proposes a uniform way to manage both changes and versioning in a same objects base or DW,
- it permits extensibility and the reusability of the different rules and strategies of a graph class evolution.